\documentclass[aps,pre,twocolumn,superscriptaddress,showpacs]{revtex4}

\usepackage{graphicx}
\usepackage{amsmath}
\usepackage{bm}

\begin{document}

\title{Tortuosity--porosity relation in the porous media flow}



\author{Maciej Matyka}
\affiliation{Institute of Theoretical Physics, University of Wroc{\l}aw, pl.\
M.\ Borna 9, 50-204 Wroc{\l}aw, Poland}

\author{Arzhang Khalili}
\affiliation{Max Planck Institute for Marine Microbiology, Celsiusstrasse 1,
D-28359 Bremen, Germany }
\affiliation{Jacobs University Bremen, Campus
Ring 1, D-28759 Bremen, Germany }

\author{Zbigniew Koza}
\affiliation{Institute of Theoretical Physics, University of Wroc{\l}aw, pl.\
M.\ Borna 9, 50-204 Wroc{\l}aw, Poland}


\date{\today}

\begin{abstract}
We study numerically the tortuosity--porosity relation in a
microscopic model of a porous medium arranged as a collectin
of freely overlapping squares.
It is demonstrated that the finite-size effects and the discretization errors, which were ignored in previous studies,
may cause significant underestimation of tortuosity.
The simple tortuosity calculation method proposed here
eliminates the need for using complicated, weighted averages. The numerical results presented here are in good agreement with an empirical relation between tortuosity ($T$)
and porosity ($\phi$) given by  $T-1\propto \ln\phi$, that was found  by others experimentally in granule packings and sediments. This relation can be also written as
$T-1\propto R S/\phi$ with $R$ and $S$ denoting  the hydraulic radius of granules and the specific surface area, respectively.
\end{abstract}



\maketitle

\section{Introduction}

In the low Reynolds number regime, flow through a porous matrix is governed by
Darcy's law that links the fluid flux (discharge per unit area) ${\bm q}$ with
the applied pressure gradient $\bm \nabla P$ by the linear relation
\begin{equation}\label{eq:darcy}
    \bm{q} = - \frac{k}{\mu} \bm\nabla P,
\end{equation}
where $\mu$ is the dynamic viscosity of the fluid and $k$ is a proportionality
constant known as  permeability \cite{Bear72}. To a large extent, the proper
description of the fluid flow through a porous medium depends on precise
relations between the physical properties involved such as permeability and
porosity ($\phi$). In particular, much attention has been paid to deriving
relations between $k$  and $\phi$ \cite{Koponen98}. In 1927 Kozeny developed a
simple capillary model for a porous medium, and proposed the relation
\begin{equation}
  \label{eq:Kozeny}
     k = c_0\frac{\phi^3}{S^2},
\end{equation}
where $S$ is the specific surface area and $c_0$ is a dimensionless Kozeny
constant that depends on the channel geometry \cite{Bear72}. Unfortunately,
Kozeny's formula is not universal and  does not hold for complicated porous
geometries \cite{Bear72}. For example, it does not take into account pore
connectivity and the fact that the specific surface area can be increased to an
arbitrarily large value by removing some of the material to roughen the porous
matrix surface in a fractal-like manner. On a purely physical ground namely,
one would expect that removal of the material from a porous matrix would
increase its permeability, whereas Kozeny's formula predicts just the opposite
\cite{Heijs95}.

One of the most widely accepted attempts to generalize
relation (\ref{eq:Kozeny}) was
proposed by Carman \cite{Carman37,Bear72,Clennell97},
who noticed that the
streamlines in a porous medium are far from being completely straight and parallel
to each other. This effect can be described by a dimensionless parameter $\ensuremath{T}$
called \emph{hydraulic tortuosity},
\begin{equation}
   \label{eq:hydrtort}
     \ensuremath{T} = \frac{\langle \lambda \rangle }{L},
\end{equation}
where $\langle \lambda \rangle$ is the average length of the fluid paths and $L$ is the geometrical length of the sample \cite{Bear72}.
Using the tortuosity, Kozeny's relation (\ref{eq:hydrtort}) can be generalized to
\begin{equation}
  \label{eq:phi1}
    k = c_0\frac{\phi^3}{T^2S^2}.
\end{equation}
By fitting experimental data, Carman concluded that $T^2$ is a constant factor
($\approx 5/2$) over a wide range of porosities. Later it was
found that $T^2$ does vary with $\phi$, and can be as
large as 50 for low porosity media
\cite{Knackstedt94,Boudreau96}.

Furthermore, it was realized that elongation of streamlines
not only affects the hydraulic discharge, but also mediates
other types of transport phenomena in the porous medium.
This resulted in introducing several distinctive, experimentally measurable tortuosities obtained from a particular transport process,
leading to diffusive \cite{Nakashima04,Garrouch01},
electrical \cite{Lorenz61,Johnson82,Garrouch01} and acoustic
\cite{Johnson82} tortuosity definitions.
There were also further  theoretical attempts to define
tortuosity \cite{Bear72,Koponen96,Clennell97}.
However, all these tortuosities, in general, differ from each other.
Except for some very simple models
\cite{Clennell97,Knackstedt94,Zhang95}, there is no clear
consensus on the relation between these definitions.
Among all these definitions, the one
expressed in Eq.\ (\ref{eq:hydrtort}) is not only the simplest, but also widely adopted in theoretical studies, for it ties tortuosity with the underlying geometry and topology of the porous medium.

It has been known since long that flow through a porous medium depends on  many
factors  such as porosity, tortuosity, granule shape and size distribution,
saturation, Reynolds number, etc. For proper understanding of transport
phenomena in porous media, however, it is essential to depart from  simple
systems with a limited number of well-defined control parameters. Therefore, in
this paper we investigate the hydraulic tortuosity (as defined in Eq.\
(\ref{eq:hydrtort})) in a creeping flow through a porous region constructed by a
two-dimensional lattice system with a uniform, randomly distributed and freely
overlapping solid squares. This model, first used by Koponen \emph{et al.}\
\cite{Koponen96}, is simple enough to allow a numerical solution with the
advantage of having porosity ($\phi$) as the only control parameter. It is
particularly suitable for studying tortuosity-porosity relation,  as at high
porosities the streamlines are almost straight,
whereas for low porosities they become wiggly.

Using the Lattice Gas Automata (LGA) method,
Koponen \emph{et al.}\ \cite{Koponen96}
 solved the flow equations for  a
porosity range of $\phi\in[0.5,1]$, and concluded that
\begin{equation}
  \label{eq:tauphi96}
     \ensuremath{T} = p(1 - \phi) + 1,
\end{equation}
where $p$ is a fitting parameter. However, later they found the relation not being consistent with the results obtained for the
porosity range $\phi\in[0.4,0.5]$, and suggested \cite{Koponen97}
to replace  relation (\ref{eq:tauphi96})  by
\begin{equation}
  \label{eq:tauphi97}
    \ensuremath{T}  =  p \frac{(1-\phi)}{(\phi-\phi_c)^m}+1,
\end{equation}
in which $\phi_c\approx 0.33$ is the percolation threshold while $p$ and $m$
are some empirical parameters. Still, as an \emph{ad hoc} formula with two
fitting parameters, this relation can not be considered  a universal law. In
addition, the data used to derive (\ref{eq:tauphi97}) suffer from systematic
errors, as neither the impact of a finite system size nor the effect of the
space discretization were taken into account.

The aim of this paper is to cary out a numerical simulation for analysing the tortuosity---porosity relation in a system of freely overlapping rectangles with a high accuracy.
In addition, we provide a simplified algorithm for  $\ensuremath{T}$ calculation without the need for implementing complicated, weighted averages of streamline lengths.

The structure of the paper is as follows.
Section \ref{sec:Model} specifies the model and the numerical techniques  used.
Special attention is paid to the description of the non-trivial numerical technique  for the
tortuosity.
Main results, including a detailed
numerical error and finite-size analysis are provided
in Sec.\ \ref{sec:results}.
Finally, the results  are discussed
in  Sec.\ \ref{sec:discussion}.


\section{Model    \label{sec:Model}}

\subsection{General description}

The system of interest consists of a square lattice ($L\times L$) in which a
number of solid squares ($a\times a$ lattice units) have been placed at random
locations to form a porous matrix ($1 \le a \ll L$). The squares are fixed in
space but free to overlap.  The only restriction is that their sides must
coincide with the underlying lattice. The remaining, void space is filled with
a liquid. The constant, external force imposed on the porous medium is aligned
with the $y$-axis of an $x$-$y$ Cartesian coordinate system to model the
gravity. Following previous works \cite{Koponen96,Koponen97}, periodic boundary
conditions have been imposed in both directions.

\begin{figure}
  \centering
\begin{tabular}{cc}
\includegraphics[scale=0.28]{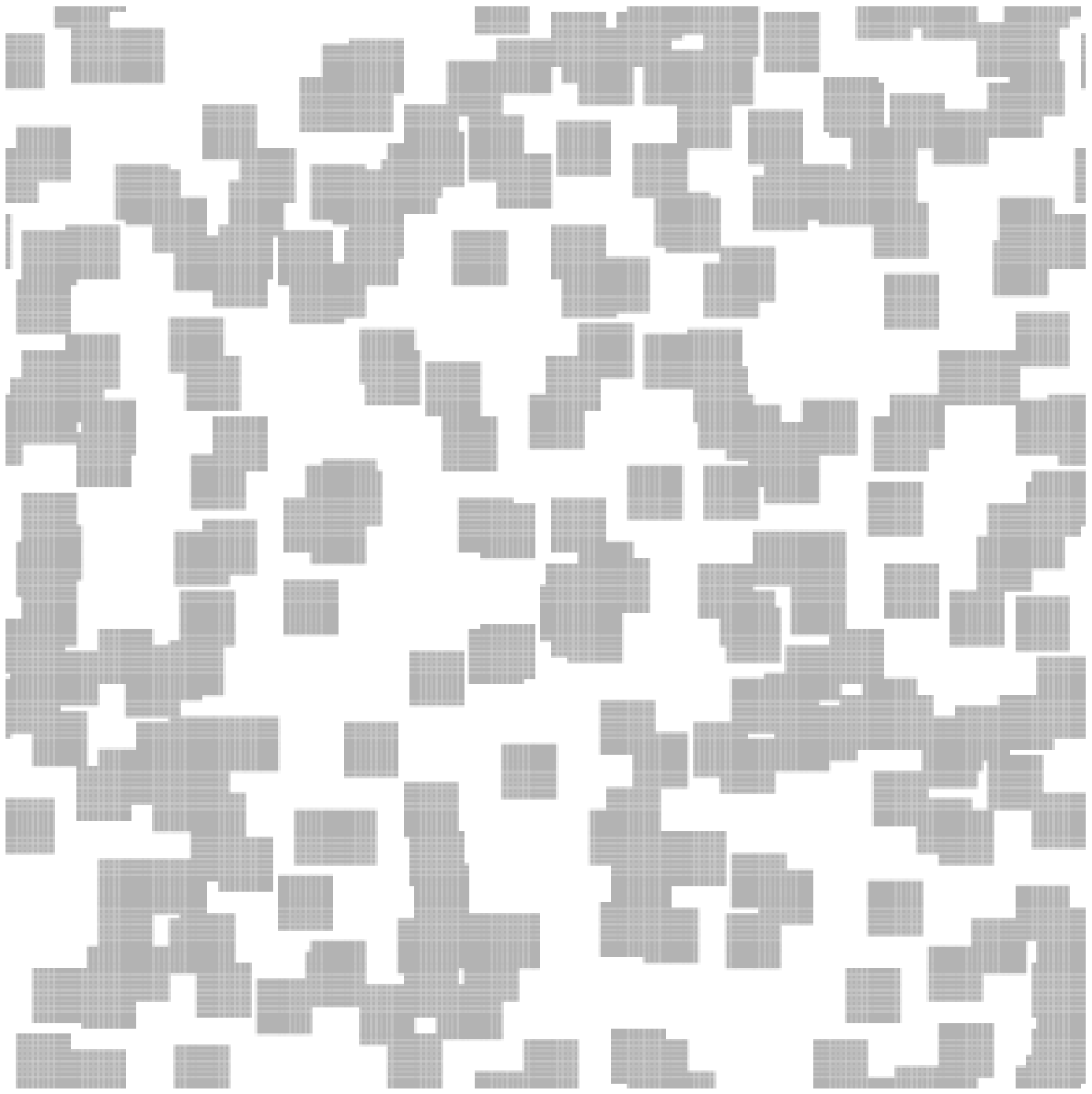}
  \hspace{0.1cm}
    &
  \hspace{0.1cm}
\includegraphics[scale=0.28]{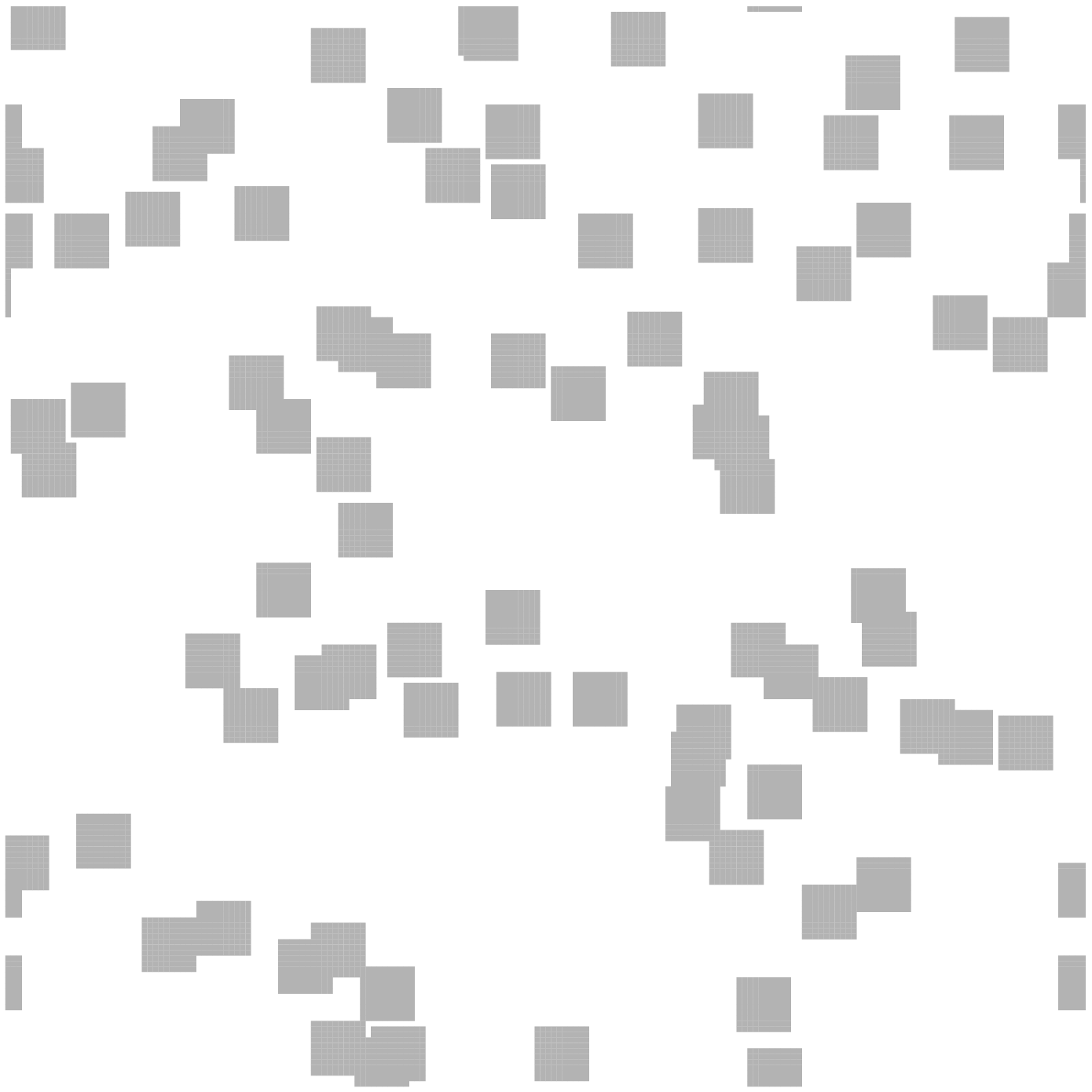} \\
a) $\phi=0.5$ \hspace{0.1cm}&\hspace{0.1cm} b) $\phi=0.8$ \\
\end{tabular}
\caption{An example of  two $800\times 800$ porous matrices
constructed by randomly placed and freely overlapping
rectangles of size $10\times 10$ for two
different porosities $\phi$.\label{pic:geometry}}
\end{figure}

Two examples of such porous systems are depicted in Fig.\
\ref{pic:geometry}. The dark areas represent fixed solid obstacles, while the white part is occupied by  the fluid.

\subsection{Numerical techniques
\label{sec:numerics}}

Numerical solution of the model defined above  consists of five main steps:
(i)    generation of a porous matrix of a known porosity;
(ii) solving the flow equations in the low Reynolds
     number regime;
(iii) finding the flow streamlines;
 (iv)   determining the tortuosity of the flow;
 (v) error analysis.

\subsubsection{Construction of the porous matrix}
A porous matrix of a given porosity $\phi$ can be generated using the method of
\cite{Koponen96,Koponen97}. Starting from an empty system,
solid squares are added at random positions until the desired
porosity has been reached.
The porosity is calculated as the fraction of unoccupied
lattice nodes.

\subsubsection{Lattice Boltzmann method for solving flow equations}

To solve the flow equations, we applied the Lattice
Boltzmann Model (LBM) \cite{Succi01} with
a single relaxation time collision operator
\cite{Bhatnagar54}. This method proved useful in
microscopic model simulations of flow through porous media
for various conditions and flow regimes
\cite{Koponen98,Pan01}.
It is a numerical technique that rests on the Boltzmann
transport equation discretized both in time and space, and
is expressed in terms of the velocity distribution
functions $n_i$ in the form
\begin{equation}
  \label{eq:lbm}
    n_i^{t+1}(\bm{r}+\bm{c}_i) = n_i^{t}(\bm{r}) + \Delta_i^{t}(\bm{r}) + F_i,
\end{equation}
where $i = 0,\ldots,8$ identifies lattice vectors $\bm{c}_i$, $t$ is an integer time step, $\bm{r}$ denotes a lattice node, $\Delta_i^t(\bm r)$ is the
collision operator at $\bm r$, and $F_i$ represents the $i$-th component of the external force.
We used the time unit equal to the relaxation time, which
yields the kinematic viscosity $\nu=1/6$ \cite{Succi01}.
This, in turn, simplifies the form of the collision operator
$\Delta_i^t(\bm r) = n_i^{\mathrm{eq}}(\bm{r}) - n^t_i(\bm r)$ with $n_i^{\mathrm{eq}}(\bm{r})$ being
the equilibrium value of $n_i$ at $\bm{r}$.
The external force was taken into
account using a method of Ref.\ \cite{Guo02}: half of the momentum was transferred directly into the equilibrium distribution
function during the collision step, whereas the other half was included into the transport equation.
Because we were interested in the solution of a slow, laminar flow, we utilized the  equilibrium distribution function
$n^{\mathrm{eq}}_i$ linearized in the velocity as
\begin{equation}
   \label{eq:distribution}
      n^{\mathrm{eq}}_i =
        w_i\rho  \left[
                          1 + 3(\bm{u}\cdot\bm{c}_i)
                \right],
\end{equation}
in which $\bm{u}$ is the macroscopic velocity vector
and $w_i$ are some weighting coefficients that
depend on the lattice structure and dimension \cite{Verberg99,HeLuo97}.

One problem with the LBM method is that it is incapable of resolving the macroscopic Navier-Stokes equations for
channels narrower than about 4 lattice
units \cite{Succi01}. This limitation becomes
particularly important at low porosities, for which
the number of very narrow passages increase enormously.
To bypass this problem, a  standard numerical mesh refinement procedure was used.
Starting from the original lattice taken to generate
the porous matrix,  each of its $L^2$ elementary quads
were subdivided into $\ensuremath{k_\mathrm{ref}}\times \ensuremath{k_\mathrm{ref}}$ smaller
quads with $\ensuremath{k_\mathrm{ref}} = 1,2,\ldots$ being the refinement level.
The resulting $\ensuremath{k_\mathrm{ref}} L\times \ensuremath{k_\mathrm{ref}} L$ computational grid of
vectors $\bm r$ in Eq.\ (\ref{eq:lbm}) will then be
formed from the centers of the small quads.
With this choice,  the identification of the interface
between the porous matrix and the free space is facilitated.

\begin{figure}
  \centering
    \includegraphics[width=0.95\columnwidth,
    clip=true]{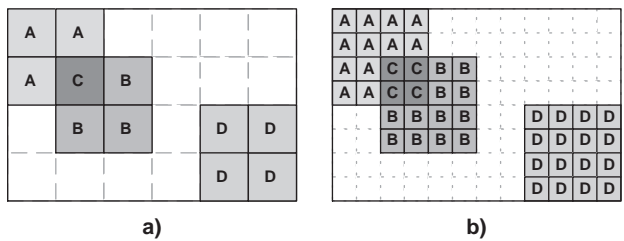}
    \caption{A schematic view of the system before
    (a) and after (b) grid refinement.
             In this example the system
             size is $4\times 6$, the porous matrix
             is made of three quads of size $2\times 2$
             (A, B and D), and the refinement level $\ensuremath{k_\mathrm{ref}}=2$.
    \label{pic:refinement}}
\end{figure}

A schematic picture of the refinement procedure is shown in Fig.\
\ref{pic:refinement}. Note that the refinement effectively increases the number of the lattice nodes between any two points
by the factor $\ensuremath{k_\mathrm{ref}}$, and that the smallest channel width
is $\ensuremath{k_\mathrm{ref}}+ 1$ lattice units.

\begin{figure}
 \centering
 \includegraphics[width=0.95\columnwidth, 
 bb="0pt 0pt 231pt 233pt"]{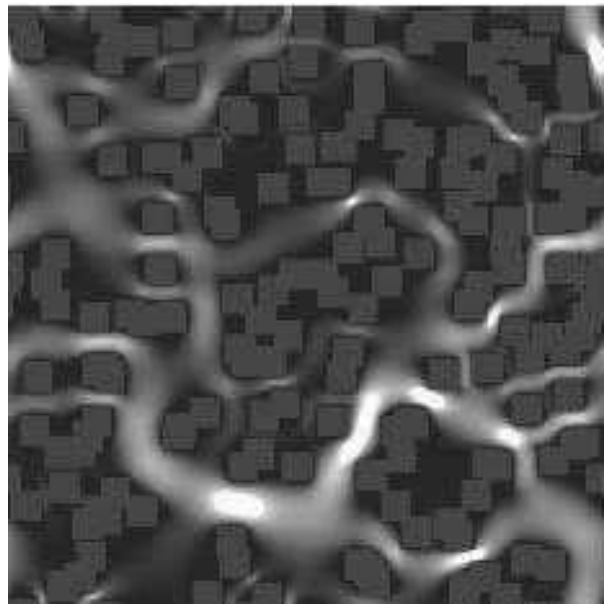}
   \caption{Velocity magnitudes squared ($u^2 = {u_x^2+v_y^2}$)
            calculated on a $600\times600$ numerical
        grids, which corresponds to a $200\times200$
            lattice (refinement level $\ensuremath{k_\mathrm{ref}} = 3$).
        The square block sizes were
                $10\times10$ (i.e., $30\times30$ after refinement)
        and the porosity was $\phi = 0.65$.
        Periodic boundary conditions were assumed in
                both directions. The grey squares in the figure represent
        the solid part of the medium, whereas the black region
        is the porespace open to fluid flow.
   \label{pic:magn1}}
\end{figure}

After initialization, the LBM computational loop of advection and collision continued for $5\times 10^3$ time steps. By using the mid-grid bounce-back rule applied to the no-slip boundaries,
second order accurate solutions, both in space and time, could be achieved \cite{Succi01}.
An example of the velocity field calculated with this method for a low-porosity medium is shown
in Fig.\ \ref{pic:magn1}.

\subsubsection{Flow streamlines}

After obtaining the velocity field $\bm{u}$ at each  grid point,
streamlines could be obtained by  solving the equation of motion for the trajectories $\bm{r}(t)$ of  massless particles
 \cite{Harlow65},
\begin{equation}
 \label{eq:MaselessMotion}
  \frac{d\bm{r}}{dt} = \bm{u}(\bm{r}).
\end{equation}
The  $\bm{u}(\bm{r})$ values of points lying between two grid nodes were obtained  using
bilinear interpolation. Due to complex boundary conditions and extreme velocity differences on the grid, the $4$-th order Runge-Kutta algorithm with adaptive time stepping was used \cite{Press86}.

\subsubsection{Tortousity}
The tortuosity is defined by Eq.\
(\ref{eq:hydrtort})  as the ratio of the average length of
all particle path lines passing through a given cross-section
during a unit time period to the width of the sample
\cite{Bear72} leading to 
\begin{equation}
  \label{eq:finaltortuosity}
     \ensuremath{T} = \frac{1}{L}
              \frac{ \int_A u_y(x) \lambda(x) dx }{ \int_A u_y(x) dx},
\end{equation}
in which $A$ is an arbitrary cross-section of the system parallel to the $x$ axis,
$\lambda(x)$ is the length of the streamline cutting $A$ at $x$, and $u_y(x)$
is the $y$ component of the trial particle velocity at $x$.

The integrals in (\ref{eq:finaltortuosity}) have been obtained in the literature either by  the Monte Carlo integration
\cite{Koponen96,Koponen97,Alam06} or by direct quadratures \cite{Knackstedt94}.
In the former method,  the lengths of the
streamlines passing  through randomly chosen points within the pore volume are averaged using proper weights.
In the latter method $T$ is approximated  by the relation
\begin{equation}
  \label{eq:DiscreteTortuosity}
     \ensuremath{T} \approx
       \frac{1}{L}
       \frac{ \sum_j u_y(x_j) \lambda(x_j) \Delta x_j }{
                            \sum_j u_y(x_j) \Delta x_j },
\end{equation}
where $\Delta x_j = x_{j+1} - x_j$ are discretization intervals of $A$. In
principle, both approaches should yield the same results, but both can be
easily misused. For example, some researchers used the Monte Carlo integration
with streamlines  passing through points chosen randomly from a uniform
distribution over the whole pore space \cite{Koponen96,Alam06},  some
others calculated streamlines cutting all lattice nodes \cite{Koponen97}, whereas others
recorded all streamlines crossing every lattice node along a chosen inlet plane
\cite{Knackstedt94,Zhang95}.
However, such `uniform' approaches are not coherent with the reality of  low
porosity systems,  in which transport is mostly carried out  only through  few
`conducting' channels (cf.\ Fig.~\ref{pic:magn1}). Consequently, the sums in
(\ref{eq:DiscreteTortuosity})  contain most probably many terms of practically
negligible magnitudes. To avoid this problem, we used Eq.\
(\ref{eq:DiscreteTortuosity}) with a constant-flux constraint between two
neighboring streamlines,
\begin{equation}
 \label{eq:cf-constraint}
  u_y(x_j)\Delta x_j = \mathrm{const}.
\end{equation}
With this choice, Eq.\ (\ref{eq:DiscreteTortuosity}) immediately simplifies to
\begin{equation}
 \label{eq:finalaverage}
   \ensuremath{T} \approx \frac{1}{L} \frac{1}{N} \sum_{j=1}^N \lambda(x_j),
\end{equation}
where $N$ is the number of the
streamlines generated. Note that all terms in this sum are of the same order of magnitude.

Thus, to calculate $T$,   a horizontal cross-section $A$ is chosen. Next, the coordinates of the initial points $x_j$ are determined using (\ref{eq:cf-constraint}), and the corresponding streamlines
are found  by solving
(\ref{eq:MaselessMotion}) in both directions until the solutions hit the system edges ($y=0$ and $y=L$). Finally their lengths are plugged into
(\ref{eq:finalaverage}).

\begin{figure}
  \centering
    \includegraphics[width=0.95\columnwidth,
                     bb="0pt 0pt 522pt 525pt", draft=false]{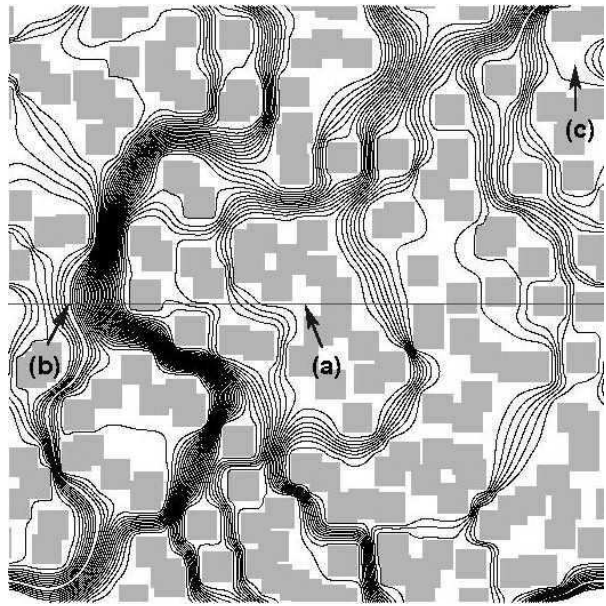}
    \caption{Streamlines generated with the constant-flux
             constraint (\protect\ref{eq:cf-constraint})
             for the same system as in Fig.\
            \protect\ref{pic:magn1} ($N=59$).
             The horizontal line represents the
             cross-section $y =L/2$
             on which the initial points $x_j$ were chosen.
             Only those streamlines are shown for which
        $\lambda(x_j)$ could be determined numerically.
              The arrows point at regions discussed in the text.
             \label{pic:str1}}
\end{figure}

%
It should be noted that not all streamlines
passing through $x_j$ cut both horizontal
edges $y=0$ and $y=L$ (see Fig.~\ref{pic:str1}).
This may happen if a streamline passes through a region
with extremely low fluid velocity. There are two types of such regions:
dead-end pores  and volumes where the fluid stream is split by an obstacle (or merges behind it).
An example of $x_j$ located in a dead-end pore is shown by arrow (a) in Fig.~\ref{pic:str1}. Arrow (b) points at $x_j$ corresponding to an incomplete streamline that flows from a region (c) where the stream merges after being split by an obstacle.

To bypass this problem, in calculation of the sum
(\ref{eq:finalaverage}) only those streamlines were taken into account,
whose lengths $\lambda_j$ can be determined.
The error induced by this procedure is discussed in
Sec.\ \ref{sec:results}.


%

\subsubsection{Error analysis}

Tortuosity values ($\ensuremath{T}$) calculated  directly from (\ref{eq:finalaverage}) contain errors arising from different sources.
While statistical errors result from randomness in the
porous matrix, discretization errors appear when approximating
the integrals in (\ref{eq:finaltortuosity}) by finite sums, and when  solving flow equations by discrete lattices. Finite-size errors could emerge also as a
consequence of approximating a macroscopic system with a microscopic model.
Details of the error analysis are addressed  in the next section.


\section{Results
  \label{sec:results}}

\begin{figure}
  \centering
    \includegraphics[width=\columnwidth]{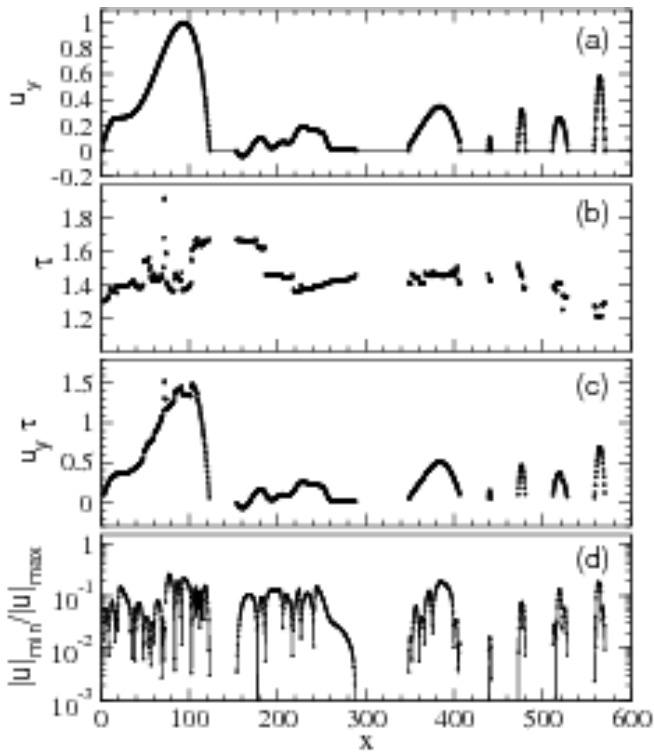}
  \caption{The basic quantities that affect tortuosity calculation         by Eq.\ (\protect\ref{eq:finaltortuosity}):
  (a) velocity component $u_y(x)$;
  (b) local tortuosity $\tau(x)=\lambda(x)/L$;
  (c) the product $u_y(x) \tau(x)$;
  (d) the ratio of the minimum to maximum speeds along streamlines.
        All quantities were determined for
        the system shown in Fig.~\protect\ref{pic:str1}         with the cross-section $y=L/2$ and the
        uniform discretization
        $\Delta x_j = L/N$ ($N=1200$).
        The tortuosity for this system
  is $T\approx1.45$.
  \label{fig:zk1}}
\end{figure}

To begin te discussion, the structure of the integrands in
(\ref{eq:finaltortuosity}) is examined.
Figure \ref{fig:zk1} shows $u_y(x)$, the local
tortuosity $\tau(x) = \lambda(x)/L$, their product
$u_y(x)\tau(x)$, and the ratio of the minimum to maximum trial particle speeds for the same system as in
Figs.\ \ref{pic:magn1} and \ref{pic:str1}.
The data for this figure were
obtained from the streamlines originating at the cross-section $y=L/2$. As
expected, the velocity profile is continuous and piecewise-differentiable, and  partially resembling that of a Poiseuille (parabolic) flow. The negative value of $u_y$ near $x=160$ indicates that some streamlines are cutting the initial cross-section many times.
This effect must be taken into account to avoid
multiple counts of the same streamline in
Eq.~(\ref{eq:finalaverage}).

In contrast to $u_y$, the local tortuosity $\tau$ is a discontinuous function of $x$ (Fig.\ \ref{fig:zk1}b).
Each jump in the  $\tau(x)$ plot
corresponds to the stream flow splitting into
(or merging from) two parts upon
meeting an obstacle.
Thus, for a finite-size system,
$\tau(x)$ is a
piece-wise continuous function with a certain
number of discontinuities, equal to the number of
,,islands'' existing in the porous domain.
Consequently,
the product $u_y(x)\tau(x)$ is
also discontinuous (Fig.~\ref{fig:zk1}c).
Moreover, the problem of finding the
coordinates of discontinuity points is
numerically ill-conditioned. These two
factors greatly complicate the determination of
the enumerator in Eq.\
(\ref{eq:finaltortuosity}), and introduce an additional
source of errors in (\ref{eq:finalaverage}).
For $x_j$ near a discontinuity point, even
small numerical errors may result in a
significant jump in $\lambda(x_j)$.
Two countermeasures were taken to reduce the impact
of this phenomenon, which is closely
related to the problem of ``missing streamlines''
discussed in Sec.\
\ref{sec:numerics}.
First, a check is made to find out how  $T$ calculated from
(\ref{eq:finalaverage}) varies with $N$. Here, an optimal value
of $N\approx L$ was found.
Second, the tortuosity was always calculated as an average over
8 different cross-sections.
This approach not only reduced the error resulting
from approximating (\ref{eq:finaltortuosity}) by (\ref{eq:finalaverage}), but
also gave some estimation on its magnitude.
The errors were found to be maximum for
low porosities, but even for $\phi = 0.45$, the relative error
was less than $0.5\%$.

The large number of discontinuities in $\tau(x)$ implies  that the fluid
velocity along a typical streamline may vary by many orders of magnitude. This
is shown  in Fig.~\ref{fig:zk1}d, in which  the ratio of the minimum to maximum
fluid speed ($u_\mathrm{min}/u_\mathrm{max}$) along the streamline cutting the
cross-section $y=L/2$ at $x$ is plotted. In this particular case
$u_\mathrm{min}/u_\mathrm{max} \leq 0.26$ and drops to 0 at all positions where
$\tau(x)$ became discontinuous.
A comparison of panels (a) and
(d) in Figure \ref{fig:zk1} shows that streamlines
passing through a region with relatively high fluid velocity
will likely hit regions where the
fluid is almost stagnant. For this reason it is essential that
Eq.~(\ref{eq:MaselessMotion}) be solved with a numerical method that uses variable step lengths and local error control.

\begin{figure}
  \centering
    \includegraphics[width=0.95\columnwidth]{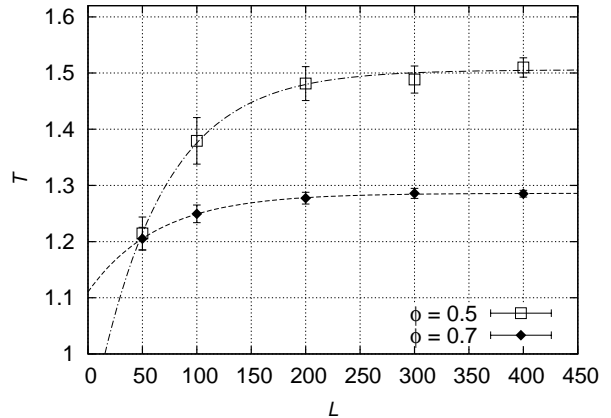}
       \caption{Tortuosity $\ensuremath{T}$ as a function of the
        system size
                $L$ for $a=10$ and $\phi=0.7$ and $0.5$,
        averaged over $M=30$
                samples. The solid line is the best fit
                calculated with Eq.\ (\ref{eq:fit}).
                The error bars represent the standard
        error of the mean.
       \label{fig:ref1}}
\end{figure}

Next, finite-size effects were analysed. Figure \ref{fig:ref1}
shows the dependency of the tortuosity $\ensuremath{T}$ on the system
size $L$ for two porosities $\phi = 0.7$ and $\phi = 0.5$
averaged over $M=30$ samples. The lines represent fits to
\begin{equation}
  \label{eq:fit}
   \ensuremath{T}(L) = \ensuremath{T}_{\infty} - b\exp(-cL),
\end{equation}
where $\ensuremath{T}_\infty$, $b$ and $c$ are free parameters.
These, as well as
many other fits, not shown here, proved to be suitable enough to
estimate the tortuosity of an infinite system
$T_\infty$. This procedure also
enabled us to estimate a characteristic system
length $L^*$ above which $T$
does not change significantly with $L$.
It turned out that $L^*\approx 200$ for
$\phi \geq 0.55$ and $L^* \approx 300$ for $0.45 \leq \phi < 0.55$. In all cases analysed, $T$ was found to be an increasing function of $L$. Thus, it becomes clear that ignoring
finite-size effects and using $L<L^*$ would lead to an
underestimation of~$T$.

\begin{figure}
  \centering
     \includegraphics[width=0.95\columnwidth]{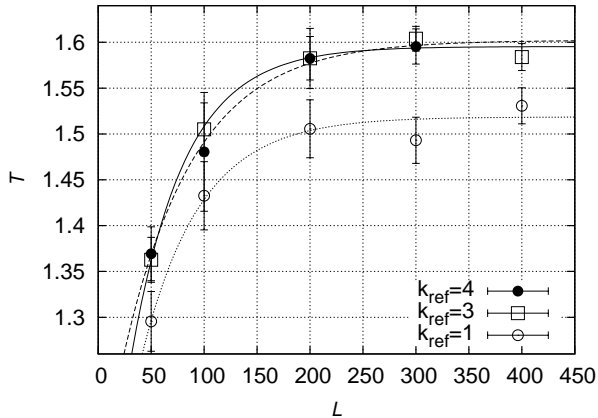}
        \caption{The dependency of $T$ on the system size
                 $L$ for
                 $\phi=0.45$ and three refinement
        levels $k_{\mathrm{ref}}$ (symbols).
                 The lines represent best fits
        to the function given by Eq.\
                 (\ref{eq:fit}).
        \label{fig:phi045}}
\end{figure}

Following this, the sensitivity of tortuosity
measurements to the numerical mesh
refinement was examined. Figure \ref{fig:phi045} depicts the values of $\ensuremath{T}(L)$ for three
refinement levels $\ensuremath{k_\mathrm{ref}} = 1,3,4$.
We found that $\ensuremath{T}_\infty$ significantly depends on
$\ensuremath{k_\mathrm{ref}}$ only for $\ensuremath{k_\mathrm{ref}} \le 3$.
This threshold value is in accord with the
criteria, mentioned above, for the
LBM method to reconstruct the Navier-Stokes equations
\cite{Succi01}.
Note that ignoring numerical mesh refinement would lead to a significant underestimation
of  $T$. This shows that narrow passages are a
relevant factor that affects
transport properties of a porous medium.
Numerical mesh refinement is thus
particularly important at low porosities,
where narrow passages are common.

\begin{figure}
  \centering
    \includegraphics[width=0.95\columnwidth]{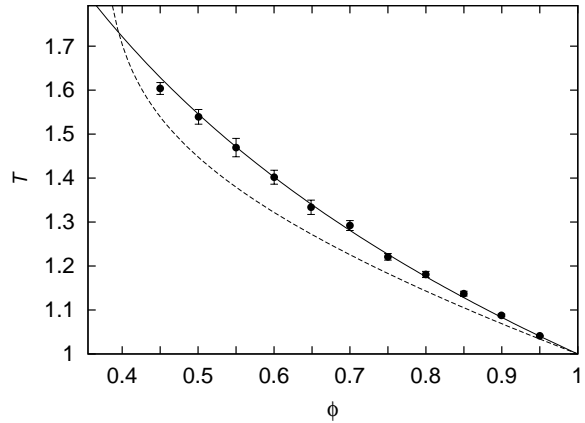}
    \caption{The dependency of the tortuosity
    $\ensuremath{T}$ on the porosity $\phi$. Our
             data obtained with the LBM method and Eq.\
        (\ref{eq:finalaverage})
            (circles);
        relation (\protect\ref{eq:tauphi97})
        derived for the same system by Kopponen
        \emph{et al.} (dashed line); the best fit to Eq.\
            (\protect\ref{fit:comiti})
            (solid line).
       \label{fig:fin}}
\end{figure}

After finding the minimal requirements on the mesh refinement
level $\ensuremath{k_\mathrm{ref}}$ and the  system size $L^*$, the
tortuosity-porosity relation could be determined.
For a given $\phi = 0.45,0.5,\ldots,0.95$,
a system size of $L= L^*$ and a refinement level of $\ensuremath{k_\mathrm{ref}} = 3$ were chosen.
At each $\phi$, $T$ was calculated for $M$ porous
matrices, with $M$ ranging from 30 (for $\phi=0.95$) to $140$
(for $\phi=0.45$), and the results are shown as circle symbols
in  Fig.\ \ref{fig:fin}. For comparison, we
also plotted the best-fit curves obtained for exactly the same system by Koponen \emph{et al.} \cite{Koponen97},
see Eq.\ (\ref{eq:tauphi97}). Obviously, the
results of Koponen \emph{et al.} lie significantly below those obtained by us.
This is due to the fact that in the work of  Koponen \emph{et al.},
a rather small system ($L  < L^*)$ is considered  without numerical mesh refinements. Hence, the discrepancy can be explained as a consequence of finite-size effects and discretization error analysis, which are missing in their study.

We fited our data to four tortuosity-porosity relations proposed by other
researchers:
\begin{subequations}
\begin{eqnarray}
  T(\phi) &=& \phi^{-p},                         \label{fit:archie}   \\
  T(\phi) &=& 1-p\ln \phi,                    \label{fit:comiti}   \\
  T(\phi) &=& 1 + p(1-\phi),                  \label{fit:linear}   \\
  T(\phi) &=& \left[1 + p(1-\phi) \right]^2,  \label{fit:boudreau} 
\end{eqnarray}
\end{subequations}
where $p$ is a parameter.
The first of them was proposed for the electric tortuosity by Archie
(1942) \cite{Archie42}. The second equation (with $p=1/2$) was found in  a
theoretical study on diffusivity of a model porous system  composed of freely
overlapping spheres by Weissberg (1966) \cite{Weissberg63}. The same relation
(with $p\approx 0.86$ and $p\approx 1.66$) was also  reported in measurements
of the hydraulic tortuosity for fixed beds of parallelepipedal particles with
different thickness-to-side ratios by Comiti and Renaud (1989) \cite{Comiti89}
and in recent measurements of electrical tortuosity in fixed beds and suspensions
of glass spheres by Barrande \emph{et al} (2007)  \cite{Barrande07}.
 Equation (\ref{fit:linear}) is an
empirical relation found for sandy ($p=2$) or clay-silt ($p=3$) sediments by
Iversen and J{\o}rgensen (1993) \cite{Iversen93}. Finally, equation
(\ref{fit:boudreau}), with $p= 32/9\pi \approx 1.1$, was recently obtained in a
model of the diffusive tortuosity in marine muds by Boudreau and Meysman (2006)
\cite{Boudreau06}.
%
%
%

Even though we treated $p$ in all these formulas as an adjustable parameter,
only Eq.\ (\ref{fit:comiti}) gave a satisfactory fit for $p = 0.80 \pm 0.01$
(the reduced chi-square statistics $\approx 0.8$), a value
 comparable with those of Comiti and Renaud
 \cite{Comiti89} ($p\approx 0.86$ and $p\approx 1.66$).
This fit is plotted in Fig.\ \ref{fig:fin} as a solid line.

The fact that our system obeys Eq.\ (\ref{fit:comiti}) has a rather interesting
and unexpected consequence. As shown previously \cite{Koponen97}, the specific
surface area $\mathrm{S}$ in the model of freely overlapping squares of side
$a$ satisfies the relation
\begin{equation}
  \label{eq:specific}
   S = -\frac{2}{R} \phi\ln\phi,
\end{equation}
with $R= a/2$ denoting hydraulic radius of obstacles. With this, equation
(\ref{fit:comiti}) simplifies to
\begin{equation}
   \label{eq:lindep}
      \ensuremath{T} - 1\propto R\frac{S }{\phi}.
\end{equation}



\section{Discussion and conclusions
  \label{sec:discussion}
}

As shown by the present study, obtaining hydraulic tortuosity from numerical simulations contains  many hidden problems, which may lead to incorrect conclusions. When a fluid stream hits an obstacle, it splits and then merges, causing a discontinuity in streamlines. The bounding streamline of each obstacle separates the two splitting (or merging) streams.
The location of such streamlines is \emph{a priori} not known, and the problem of finding the streamlines within those regions is
numerically ill-conditioned.
If the system is sufficiently large, it is inevitable that majority of streamlines pass through such ``ill-conditioned'' regions. Moreover, the
velocity magnitude along the streamlines can vary by many orders of magnitude.

These problems resemble those encountered in another type of computer
simulations, molecular dynamics, where discontinuities (as well as
large variations in velocity) are caused by collisions. After several such events, the computer-generated particle trajectories have
literally nothing to do with the exact solutions. Still, molecular dynamics is one of the most successful methods of computer physics. The reason is that physical quantities never depend on exact trajectories of the individual particles---it is sufficient to ensure that the solution keeps on a constant energy surface.
This analogy makes us believe that despite all difficulties, hydraulic
tortuosity is a well-defined quantity that can be reliably obtained by
numerical methods. All simulations we performed with  different cross-sections,
different numbers of streamlines, different choices of streamline starting points,
different numerical ordinary differential equation solvers,
resulted in almost the same numerical values.
This implies that---just as in molecular dynamics---small local errors,
which are unavoidable in computer simulations, are of marginal importance.
This corresponds to the fact that in real fluid flow, trajectories of individual molecules are not limited to single ``theoretical'' streamlines, but are affected by diffusion at low velocities or turbulence at higher
velocities.

It is also interesting to mention that the problem of finding $T$ is easiest at
high porosities, where nearly each obstacle constitutes a separate island.
Although the number of discontinuities is very large, in general they tend to
average out. At low porosities, however, severe problems may arise as
discontinuities are much fewer in number (which means no ``averaging out'') but
much larger in magnitude (which results from increased island sizes). With this
study we have demonstrated how sensitive tortuosity computations are to
finite-size effects, discretization errors and large variation of fluid speed
along streamlines. The system size must be large enough to ensure development
of chaotic ``splitting and merging'' flows, that are characteristic of real
granular systems. Discretization errors creep into the system from  several
places, most notably in narrow channels, and can be avoided by numerical mesh
refinement. Our results concerning large fluid velocity variations along
streamlines are a clear indication for revising those tortuosity definitions
which assume a constant fluid velocity along a streamline
\cite{Knackstedt94,Zhang95}. They also show that numerical determination of
streamlines requires using advanced numerical integrators with adaptive step
lengths and local error control.

When streamlines are generated using the constant-flux constraint, the
tortuosity can be calculated simply as an average over the streamline lengths.
This method reduces the computation errors and does away the need for
determining the  streamlines in dead-end pores.


%
%

The numerical data presented in this study were found to be in good agreement
with those generated by Eq.\ (\ref{fit:comiti}), obtained from previous
experimental studies. However, this relation cannot be applied close to the
percolation threshold, where tortuosity diverges. Note also that Eq.
(\ref{fit:comiti}) was found to describe both hydraulic and diffusive
tortuosities, i.e.\ quantities that are certainly correlated, but in a way that
has not been well established yet. It is not clear whether our findings reflect
equivalence of these quantities, or whether they can be considered as a
coincidence.

We also found that in the model of freely overlapping squares, a very  simple
relation (\ref{eq:lindep}) holds between tortuosity, porosity and the specific
surface area. This equation is closely related to Eq.\ (\ref{fit:comiti}), and
it expected to be valid for all systems, for which Eq.\ (\ref{fit:comiti})
holds.


\begin{thebibliography}{29}
\expandafter\ifx\csname natexlab\endcsname\relax\def\natexlab#1{#1}\fi
\expandafter\ifx\csname bibnamefont\endcsname\relax
  \def\bibnamefont#1{#1}\fi
\expandafter\ifx\csname bibfnamefont\endcsname\relax
  \def\bibfnamefont#1{#1}\fi
\expandafter\ifx\csname citenamefont\endcsname\relax
  \def\citenamefont#1{#1}\fi
\expandafter\ifx\csname url\endcsname\relax
  \def\url#1{\texttt{#1}}\fi
\expandafter\ifx\csname urlprefix\endcsname\relax\def\urlprefix{URL }\fi
\providecommand{\bibinfo}[2]{#2}
\providecommand{\eprint}[2][]{\url{#2}}

\bibitem[{\citenamefont{Bear}(1972)}]{Bear72}
\bibinfo{author}{\bibfnamefont{J.}~\bibnamefont{Bear}},
  \emph{\bibinfo{title}{Dynamics of fluids in porous media}}
  (\bibinfo{publisher}{Elsevier}, \bibinfo{address}{New York},
  \bibinfo{year}{1972}).

\bibitem[{\citenamefont{Koponen et~al.}(1998)\citenamefont{Koponen, Kandhai,
  Hell\'en, Alava, Hoekstra, Kataja, Niskanen, Sloot, and Timonen}}]{Koponen98}
\bibinfo{author}{\bibfnamefont{A.}~\bibnamefont{Koponen}},
  \bibinfo{author}{\bibfnamefont{D.}~\bibnamefont{Kandhai}},
  \bibinfo{author}{\bibfnamefont{E.}~\bibnamefont{Hell\'en}},
  \bibinfo{author}{\bibfnamefont{M.}~\bibnamefont{Alava}},
  \bibinfo{author}{\bibfnamefont{A.}~\bibnamefont{Hoekstra}},
  \bibinfo{author}{\bibfnamefont{M.}~\bibnamefont{Kataja}},
  \bibinfo{author}{\bibfnamefont{K.}~\bibnamefont{Niskanen}},
  \bibinfo{author}{\bibfnamefont{P.}~\bibnamefont{Sloot}}, \bibnamefont{and}
  \bibinfo{author}{\bibfnamefont{J.}~\bibnamefont{Timonen}},
  \bibinfo{journal}{Phys. Rev. Lett.} \textbf{\bibinfo{volume}{80}},
  \bibinfo{pages}{716} (\bibinfo{year}{1998}).

\bibitem[{\citenamefont{Heijs and Lowe}(1995)}]{Heijs95}
\bibinfo{author}{\bibfnamefont{A.~W.~J.} \bibnamefont{Heijs}} \bibnamefont{and}
  \bibinfo{author}{\bibfnamefont{C.~P.} \bibnamefont{Lowe}},
  \bibinfo{journal}{Phys. Rev. E} \textbf{\bibinfo{volume}{51}},
  \bibinfo{pages}{4346} (\bibinfo{year}{1995}).

\bibitem[{\citenamefont{Carman}(1937)}]{Carman37}
\bibinfo{author}{\bibfnamefont{P.~C.} \bibnamefont{Carman}},
  \bibinfo{journal}{Trans. Inst. Chem. Eng.} \textbf{\bibinfo{volume}{15}},
  \bibinfo{pages}{150} (\bibinfo{year}{1937}).

\bibitem[{\citenamefont{Clennell}(1997)}]{Clennell97}
\bibinfo{author}{\bibfnamefont{M.~B.} \bibnamefont{Clennell}},
  \bibinfo{journal}{Geological Society, London, Special Publications}
  \textbf{\bibinfo{volume}{122}}, \bibinfo{pages}{299} (\bibinfo{year}{1997}).

\bibitem[{\citenamefont{Knackstedt and Zhang}(1994)}]{Knackstedt94}
\bibinfo{author}{\bibfnamefont{M.~A.} \bibnamefont{Knackstedt}}
  \bibnamefont{and} \bibinfo{author}{\bibfnamefont{X.}~\bibnamefont{Zhang}},
  \bibinfo{journal}{Phys. Rev. E} \textbf{\bibinfo{volume}{50}},
  \bibinfo{pages}{2134} (\bibinfo{year}{1994}).

\bibitem[{\citenamefont{Boudreau}(1996)}]{Boudreau96}
\bibinfo{author}{\bibfnamefont{B.~P.} \bibnamefont{Boudreau}},
  \bibinfo{journal}{Geochimica et Cosmochimica Acta}
  \textbf{\bibinfo{volume}{60}}, \bibinfo{pages}{3139} (\bibinfo{year}{1996}).

\bibitem[{\citenamefont{Nakashima and Yamaguchi}(2004)}]{Nakashima04}
\bibinfo{author}{\bibfnamefont{Y.}~\bibnamefont{Nakashima}} \bibnamefont{and}
  \bibinfo{author}{\bibfnamefont{T.}~\bibnamefont{Yamaguchi}},
  \bibinfo{journal}{Bulletin of the Geological Survey of Japan}
  \textbf{\bibinfo{volume}{55}}, \bibinfo{pages}{93} (\bibinfo{year}{2004}).

\bibitem[{\citenamefont{Garrouch et~al.}(2001)\citenamefont{Garrouch, Ali, and
  Qasem}}]{Garrouch01}
\bibinfo{author}{\bibfnamefont{A.~A.} \bibnamefont{Garrouch}},
  \bibinfo{author}{\bibfnamefont{L.}~\bibnamefont{Ali}}, \bibnamefont{and}
  \bibinfo{author}{\bibfnamefont{F.}~\bibnamefont{Qasem}},
  \bibinfo{journal}{Ind. Eng. Chem. Res.} \textbf{\bibinfo{volume}{40}},
  \bibinfo{pages}{4363} (\bibinfo{year}{2001}).

\bibitem[{\citenamefont{Lorenz}(1961)}]{Lorenz61}
\bibinfo{author}{\bibfnamefont{P.~B.} \bibnamefont{Lorenz}},
  \bibinfo{journal}{Nature} \textbf{\bibinfo{volume}{189}},
  \bibinfo{pages}{386} (\bibinfo{year}{1961}).

\bibitem[{\citenamefont{Johnson et~al.}(1982)\citenamefont{Johnson, Plona,
  Scala, Pasierb, and Kojima}}]{Johnson82}
\bibinfo{author}{\bibfnamefont{D.~L.} \bibnamefont{Johnson}},
  \bibinfo{author}{\bibfnamefont{T.~J.} \bibnamefont{Plona}},
  \bibinfo{author}{\bibfnamefont{C.}~\bibnamefont{Scala}},
  \bibinfo{author}{\bibfnamefont{F.}~\bibnamefont{Pasierb}}, \bibnamefont{and}
  \bibinfo{author}{\bibfnamefont{H.}~\bibnamefont{Kojima}},
  \bibinfo{journal}{Phys. Rev. Lett.} \textbf{\bibinfo{volume}{49}},
  \bibinfo{pages}{1840} (\bibinfo{year}{1982}).

\bibitem[{\citenamefont{Koponen et~al.}(1996)\citenamefont{Koponen, Kataja, and
  Timonen}}]{Koponen96}
\bibinfo{author}{\bibfnamefont{A.}~\bibnamefont{Koponen}},
  \bibinfo{author}{\bibfnamefont{M.}~\bibnamefont{Kataja}}, \bibnamefont{and}
  \bibinfo{author}{\bibfnamefont{J.}~\bibnamefont{Timonen}},
  \bibinfo{journal}{Phys. Rev. E} \textbf{\bibinfo{volume}{54}},
  \bibinfo{pages}{406} (\bibinfo{year}{1996}).

\bibitem[{\citenamefont{Zhang and Knackstedt}(1995)}]{Zhang95}
\bibinfo{author}{\bibfnamefont{X.}~\bibnamefont{Zhang}} \bibnamefont{and}
  \bibinfo{author}{\bibfnamefont{M.~A.} \bibnamefont{Knackstedt}},
  \bibinfo{journal}{Geophys. Res. Lett.} \textbf{\bibinfo{volume}{22}},
  \bibinfo{pages}{2333} (\bibinfo{year}{1995}).

\bibitem[{\citenamefont{Koponen et~al.}(1997)\citenamefont{Koponen, Kataja, and
  Timonen}}]{Koponen97}
\bibinfo{author}{\bibfnamefont{A.}~\bibnamefont{Koponen}},
  \bibinfo{author}{\bibfnamefont{M.}~\bibnamefont{Kataja}}, \bibnamefont{and}
  \bibinfo{author}{\bibfnamefont{J.}~\bibnamefont{Timonen}},
  \bibinfo{journal}{Phys. Rev. E} \textbf{\bibinfo{volume}{56}},
  \bibinfo{pages}{3319} (\bibinfo{year}{1997}).

\bibitem[{\citenamefont{Succi}(2001)}]{Succi01}
\bibinfo{author}{\bibfnamefont{S.}~\bibnamefont{Succi}},
  \emph{\bibinfo{title}{The Lattice Boltzmann Equation for Fluid Dynamics and
  Beyond}} (\bibinfo{publisher}{Clarendon Press}, \bibinfo{address}{New York},
  \bibinfo{year}{2001}).

\bibitem[{\citenamefont{Bhatnagar et~al.}(1954)\citenamefont{Bhatnagar, Gross,
  and Krook}}]{Bhatnagar54}
\bibinfo{author}{\bibfnamefont{P.~L.} \bibnamefont{Bhatnagar}},
  \bibinfo{author}{\bibfnamefont{E.~P.} \bibnamefont{Gross}}, \bibnamefont{and}
  \bibinfo{author}{\bibfnamefont{M.}~\bibnamefont{Krook}},
  \bibinfo{journal}{Phys. Rev.} \textbf{\bibinfo{volume}{94}},
  \bibinfo{pages}{511} (\bibinfo{year}{1954}).

\bibitem[{\citenamefont{{Pan} et~al.}(2001)\citenamefont{{Pan}, {Hilpert}, and
  {Miller}}}]{Pan01}
\bibinfo{author}{\bibfnamefont{C.}~\bibnamefont{{Pan}}},
  \bibinfo{author}{\bibfnamefont{M.}~\bibnamefont{{Hilpert}}},
  \bibnamefont{and} \bibinfo{author}{\bibfnamefont{C.~T.}
  \bibnamefont{{Miller}}}, \bibinfo{journal}{Phys. Rev. E}
  \textbf{\bibinfo{volume}{64}}, \bibinfo{pages}{066702}
  (\bibinfo{year}{2001}).

\bibitem[{\citenamefont{{Guo} et~al.}(2002)\citenamefont{{Guo}, {Zheng}, and
  {Shi}}}]{Guo02}
\bibinfo{author}{\bibfnamefont{Z.}~\bibnamefont{{Guo}}},
  \bibinfo{author}{\bibfnamefont{C.}~\bibnamefont{{Zheng}}}, \bibnamefont{and}
  \bibinfo{author}{\bibfnamefont{B.}~\bibnamefont{{Shi}}},
  \bibinfo{journal}{Phys. Rev. E} \textbf{\bibinfo{volume}{65}},
  \bibinfo{pages}{046308} (\bibinfo{year}{2002}).

\bibitem[{\citenamefont{{Verberg} and {Ladd}}(1999)}]{Verberg99}
\bibinfo{author}{\bibfnamefont{R.}~\bibnamefont{{Verberg}}} \bibnamefont{and}
  \bibinfo{author}{\bibfnamefont{A.~J.~C.} \bibnamefont{{Ladd}}},
  \bibinfo{journal}{Phys. Rev. E} \textbf{\bibinfo{volume}{60}},
  \bibinfo{pages}{3366} (\bibinfo{year}{1999}).

\bibitem[{\citenamefont{He and Luo}(1997)}]{HeLuo97}
\bibinfo{author}{\bibfnamefont{X.}~\bibnamefont{He}} \bibnamefont{and}
  \bibinfo{author}{\bibfnamefont{L.-S.} \bibnamefont{Luo}},
  \bibinfo{journal}{Phys. Rev. E} \textbf{\bibinfo{volume}{56}},
  \bibinfo{pages}{6811} (\bibinfo{year}{1997}).

\bibitem[{\citenamefont{Harlow and Welch}(1965)}]{Harlow65}
\bibinfo{author}{\bibfnamefont{F.~H.} \bibnamefont{Harlow}} \bibnamefont{and}
  \bibinfo{author}{\bibfnamefont{J.~E.} \bibnamefont{Welch}},
  \bibinfo{journal}{Phys. of Fluids} \textbf{\bibinfo{volume}{8}},
  \bibinfo{pages}{2182} (\bibinfo{year}{1965}).

\bibitem[{\citenamefont{Press et~al.}(1986)\citenamefont{Press, Flannery,
  Teukolsky, and Vetterling}}]{Press86}
\bibinfo{author}{\bibfnamefont{W.~H.} \bibnamefont{Press}},
  \bibinfo{author}{\bibfnamefont{B.~P.} \bibnamefont{Flannery}},
  \bibinfo{author}{\bibfnamefont{S.}~\bibnamefont{Teukolsky}},
  \bibnamefont{and} \bibinfo{author}{\bibfnamefont{W.~T.}
  \bibnamefont{Vetterling}}, \emph{\bibinfo{title}{Numerical Recipes: The Art
  of Scientific Computing}} (\bibinfo{publisher}{Cambridge University Press},
  \bibinfo{address}{Cambridge (UK) and New York}, \bibinfo{year}{1986}).

\bibitem[{\citenamefont{Alam et~al.}(2006)\citenamefont{Alam, Byholm, and
  Toivakka}}]{Alam06}
\bibinfo{author}{\bibfnamefont{P.}~\bibnamefont{Alam}},
  \bibinfo{author}{\bibfnamefont{T.}~\bibnamefont{Byholm}}, \bibnamefont{and}
  \bibinfo{author}{\bibfnamefont{M.}~\bibnamefont{Toivakka}},
  \bibinfo{journal}{Nordic Pulp and Paper Research Journal}
  \textbf{\bibinfo{volume}{21}}, \bibinfo{pages}{670} (\bibinfo{year}{2006}).

\bibitem[{\citenamefont{Archie}(1942)}]{Archie42}
\bibinfo{author}{\bibfnamefont{G.}~\bibnamefont{Archie}},
  \bibinfo{journal}{Transactions of the American Institute Mining,
  Metallurgical and Petroleum Engineers} \textbf{\bibinfo{volume}{146}},
  \bibinfo{pages}{54–} (\bibinfo{year}{1942}).

\bibitem[{\citenamefont{Weissberg}(1963)}]{Weissberg63}
\bibinfo{author}{\bibfnamefont{H.~L.} \bibnamefont{Weissberg}},
  \bibinfo{journal}{J. Appl. Phys.} \textbf{\bibinfo{volume}{34}},
  \bibinfo{pages}{2636} (\bibinfo{year}{1963}).

\bibitem[{\citenamefont{Comiti and Renaud}(1989)}]{Comiti89}
\bibinfo{author}{\bibfnamefont{J.}~\bibnamefont{Comiti}} \bibnamefont{and}
  \bibinfo{author}{\bibfnamefont{M.}~\bibnamefont{Renaud}},
  \bibinfo{journal}{Chem. Eng. Sci.} \textbf{\bibinfo{volume}{44}},
  \bibinfo{pages}{1539} (\bibinfo{year}{1989}).

\bibitem[{\citenamefont{Barrande et~al.}(2007)\citenamefont{Barrande, Bouchet,
  and Denoyel}}]{Barrande07}
\bibinfo{author}{\bibfnamefont{M.}~\bibnamefont{Barrande}},
  \bibinfo{author}{\bibfnamefont{R.}~\bibnamefont{Bouchet}}, \bibnamefont{and}
  \bibinfo{author}{\bibfnamefont{R.}~\bibnamefont{Denoyel}},
  \bibinfo{journal}{Anal. Chem.} \textbf{\bibinfo{volume}{79}},
  \bibinfo{pages}{9115–} (\bibinfo{year}{2007}).

\bibitem[{\citenamefont{Iversen and J{\o}rgensen}(1993)}]{Iversen93}
\bibinfo{author}{\bibfnamefont{N.}~\bibnamefont{Iversen}} \bibnamefont{and}
  \bibinfo{author}{\bibfnamefont{B.~B.} \bibnamefont{J{\o}rgensen}},
  \bibinfo{journal}{Geochimica et Cosmochimica} \textbf{\bibinfo{volume}{57}},
  \bibinfo{pages}{571} (\bibinfo{year}{1993}).

\bibitem[{\citenamefont{Boudreau and Meysman}(2006)}]{Boudreau06}
\bibinfo{author}{\bibfnamefont{B.~P.} \bibnamefont{Boudreau}} \bibnamefont{and}
  \bibinfo{author}{\bibfnamefont{F.~J.} \bibnamefont{Meysman}},
  \bibinfo{journal}{Geology} \textbf{\bibinfo{volume}{34}},
  \bibinfo{pages}{693–} (\bibinfo{year}{2006}).

\end{thebibliography}
\end{document}